\documentstyle[sprocl,epsfig]{article}

\def\Journal#1#2#3#4{{#1} {\bf #2}, #3 (#4)}

\def\PLB{{\em Phys. Lett.} B}

\def\PRD{{\em Phys. Rev.} D}

\newcommand{\bm}[1]{\mbox{\boldmath $#1$}}
\newcommand{\xb}{x_{\scriptscriptstyle B}}

\begin{document}
\renewcommand{\thefootnote}{\fnsymbol{footnote}}
\unitlength=1mm

\vspace*{-21mm}
\noindent
{\small
hep-ph/9805410 \hfill VUTH 98-16\\
NIKHEF 98-012 \hfill FNT/T-98/04}\\[2mm]
\title{SPIN PHYSICS WITH SPIN-0 HADRONS
\footnote{
Presented at the workshop on "Deep Inelastic Scattering and
QCD, DIS98", Brussels (1998)}
} 

\author{\underline{R.~JAKOB}}
\address{Universit\`{a} di Pavia and INFN, Sezione di Pavia,\\
Via Bassi 6, I-27100 Pavia, Italy, e-mail:jakob@pv.infn.it} 

\author{D.~BOER}
\address{National Institute for Nuclear Physics and High Energy Physics
(NIKHEF)\\P.O.Box 41882, NL-1009 DB Amsterdam, the Netherlands} 

\author{P.J.~MULDERS}

\address{Department of Physics and Astronomy, Free University,\\De Boelelaan
1081, NL-1081 HV Amsterdam, the Netherlands}

%%%%%%%%%%%%%%%%%%%%%%%%%%%%%%%%%%%%%%%%%%%%%%%%%%%%%%%%%%%%%%

\maketitle
\setcounter{footnote}{0}
\renewcommand{\thefootnote}{\alph{footnote}}
\abstracts{
We discuss various azimuthal asymmetries in semi-inclusive
DIS and $e^+e^-\to h_1h_2X$ which involve chiral odd
quantities like the transversity distribution $h_1$ and a fragmentation
function $H_1^\perp$. For the fragmentation described by $H_1^\perp$ 
azimuthal angular dependence has to be measured, but no polarization vector
of a final state hadron. We present first
results on asymmetries including the electroweak currents in one-hadron
inclusive DIS.}
%
%%%%%%%%%%%%%%%%%%%%%%%%%%%%%%%%%%%%%%%%%%%%
\section{Transversity Distribution $h_1(x)$}
Besides the quite well-known parton distribution $f_1(x)$ there are
two different spin distributions: the longitudinal spin distribution
$g_1(x)$, on which a number of new experimental 
results have been reported at this conference~\cite{spin-res}; the 
second one, the transversity distribution\footnote{The function $h_1(x)$ was 
first discussed by Ralston and Soper~\cite{ral-79} in Drell-Yan 
scattering.}, $h_1(x)$, remains completely unknown as far as
experimental data are 
concerned.\footnote{The alternative notations $q(x)$, $\Delta q(x)$ and 
$\delta q(x)$ (or sometimes $\Delta_T q(x)$ for the latter) are also in use
instead of $f_1$, $g_1$ and $h_1$, respectively.} The transversity 
distribution $h_1$ is equally important for the
description of quarks in nucleons as the more familiar function 
$g_1$; their information is complementary.\\ 
\begin{figure}[h]
\begin{center}
%\fbox{
\begin{picture}(116,12) 
\put(0,0.5){\epsfig{figure=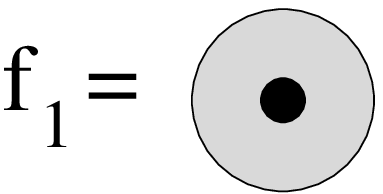,height=9mm}}
\put(28,0.5){\epsfig{figure=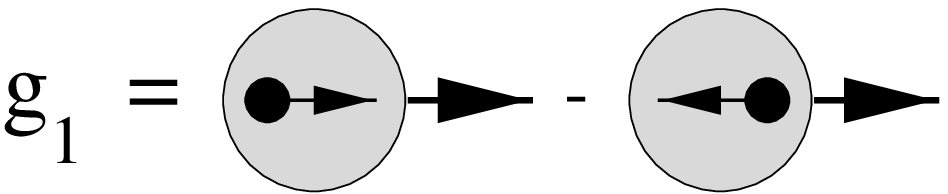,height=9mm}}
\put(78,0.5){\epsfig{figure=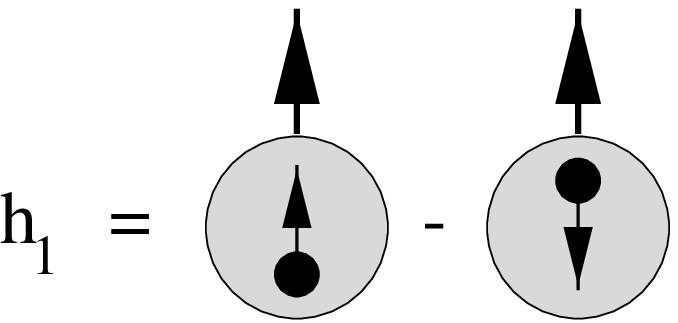,height=16.6mm}}
\end{picture}
%}
\caption{Schematic interpretation of the three distribution functions
$f_1(x)$, $g_1(x)$ and $h_1(x)$ in terms of (differences of) probabilities.
The longitudinal direction is assumed to be along the horizontal axis.}
\end{center} 
\end{figure}
The reason why $h_1$ is not determined yet,
is the fact that it is a chiral odd function, and consequently suppressed
in simple processes like totally inclusive DIS~\cite{jaf-91}. The quark 
content of $h_1(x)$ in terms of left-handed and right-handed quarks 
(defined via projection operators $P_{R/L}=(1\pm \gamma_5)/2$) is of the form 
$\bar RL-\bar LR$, i.e., contains a transition from $R$ to $L$ and 
vice versa. Since QCD interactions conserve chirality,
$h_1(x)$ cannot occur alone in a process (case $a$ in the figure below), but
has to be accompanied by a second chiral odd quantity,
like e.g.~the fragmentation function $H_1(z)$ (case $c$ below) which is the
pendant to $h_1$ for the fragmentation; obviously also the fragmentation 
function $H_1$ cannot occur alone (as in case $b$), i.e., is not accessible 
in $e^+e^-\to hX$ (see also a recent article by Jaffe~\cite{jaf-97}).
\begin{figure}[h]
\begin{center}
%\fbox{
\begin{picture}(116,30) 
\put(-2,5){\epsfig{figure=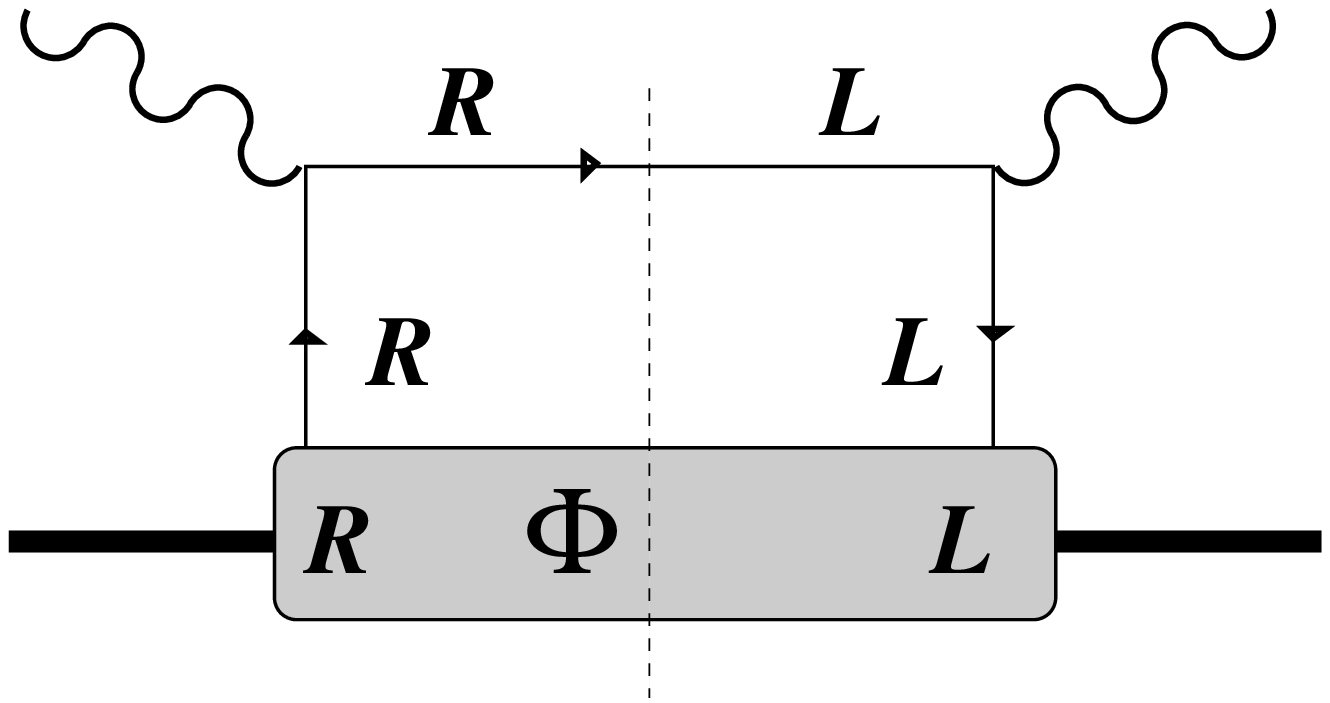,width=40mm}}
\put(41,5){\epsfig{figure=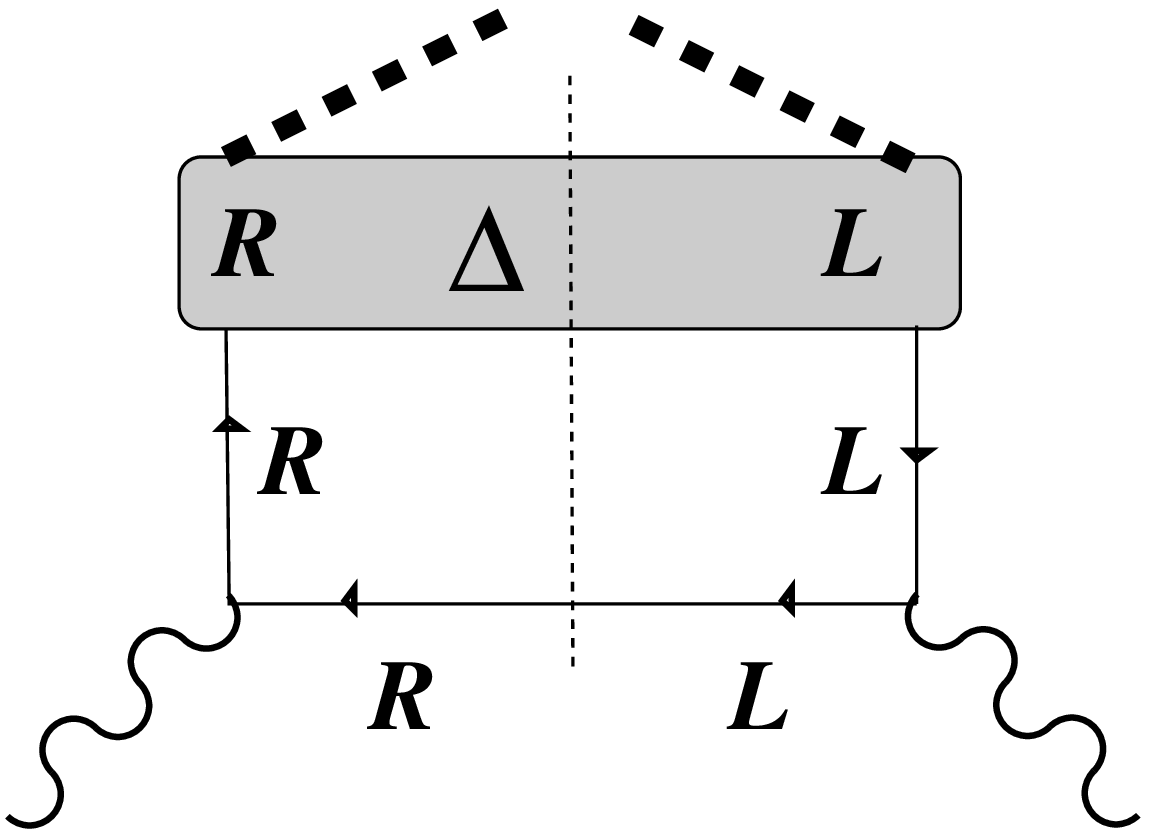,width=35mm}}
\put(82,5){\epsfig{figure=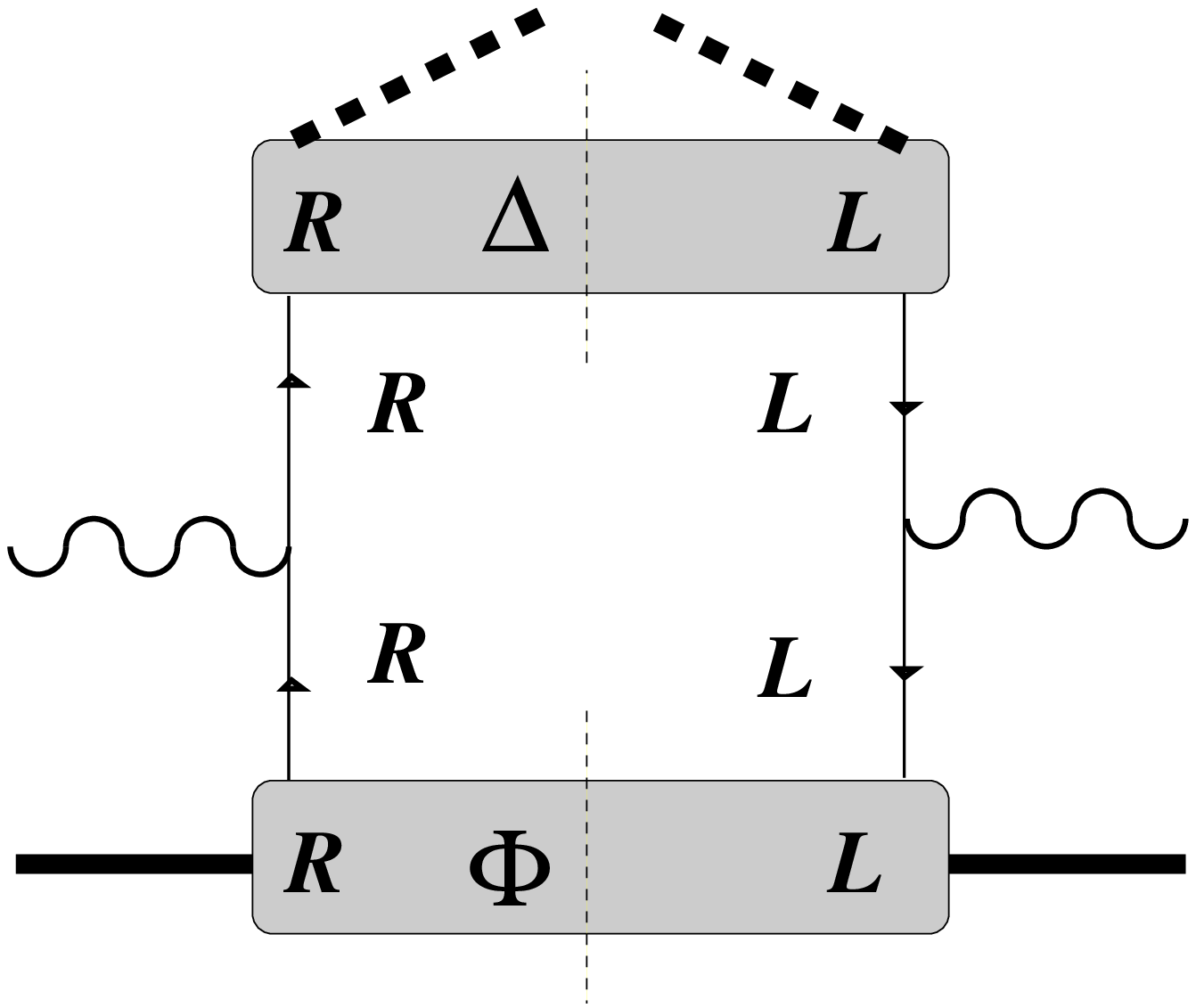,width=35mm}}
\put(16,1){$a)$}\put(56,1){$b)$}\put(97,1){$c)$}
\end{picture}
%}
\caption{Chiral odd functions can hardly be probed in the simple processes 
$a)$ (DIS) and $b)$ ($e^+e^-\to hX$), where they contribute only
via (suppressed) quark mass effects. But they are 
accessible when combined with another 
chiral odd quantity like e.g. in $c)$ ($\ell H\to\ell^\prime hX$).}
\end{center}
\end{figure}
\vspace*{-6mm}
%%%%%%%%%%%%%%%%%%%%%%%%%%%%%%%%%%%%%%%%%%%%%%%%%%%%%%%%%
\section{Azimuthal Asymmetry in SIDIS / `Collins Effect'}
To measure transversity in one-hadron inclusive DIS one can look for 
parts of the cross section involving $h_1$ and $H_1$. It has the 
disadvantage that the polarization vector of an observed 
spin-1/2 hadron in the final state has to be determined. Although not 
impossible in principle, for instance the decay of $\Lambda$'s would offer 
this possibility\cite{kot-98}, it is experimentally difficult.\\
Alternatively, one can measure azimuthal angular dependences in the
production of spin-0 or (on average) unpolarized hadrons. This production is 
described by the transverse momentum dependent 
fragmentation function\\
$H_1^\perp(x,{\bm k}_T)$ which is
also chiral odd and, moreover, ``naive time-reversal odd'', i.e., 
non-vanishing only due to final state interactions. For a more 
detailed discussion of $H_1^\perp(x,{\bm k}_T)$ see 
elsewhere~\cite{boe-97a}. The 
use of a ``naive time-reversal odd'' observable and azimuthal angular 
dependence to obtain information on $h_1(x)$ was proposed by 
Collins~\cite{col-93}. 
A particular useful reformulation of this proposal for 
$\ell H\to\ell^\prime hX$ can be given in terms of
appropriately weighted cross sections 
\begin{equation}
\left\langle W\right\rangle\;
\stackrel{\mbox{def}}{=}\;\int\,\frac{d\phi^\ell}{2\pi}\,d^2\bm{q}_T\,W\,
\frac{d\sigma(eH\to e^\prime hX)}{d\xb\,dz_h\,dy\,d\phi^\ell\,d^2\bm{q}_T}
\;\;.
\end{equation}
We find ($\phi_h^\ell$, $\phi_S^\ell$ are the azimuthal angles of $h$ and the
target spin, respectively)
\begin{equation}
\left\langle\frac{Q_T}{M_h}\sin(\phi_h^\ell+\phi_S^\ell)\right\rangle=
-\frac{4\pi\alpha^2s}{Q^4}\,|\bm{S}_T|\,(1-y)\,
\sum_{a,\bar a}\,e_a^2\,\xb\;h_1^a(\xb)\,H_1^{\perp(1)a}(z_h)
\end{equation}
containing the (flavor summed) product of the transversity distribution with
a ${\bm k}_T^2$-moment of $H_1^\perp$
\begin{equation}
H_1^{\perp(n)}(z)\;
\stackrel{\mbox{def}}{=}\;z^2\int d^2{\bm k}_T 
\left(\frac{{\bm k}_T^2}{2M_h^2}\right)^n
H_1^\perp(z,-z{\bm k}_T).
\end{equation} 
%
%%%%%%%%%%%%%%%%%%%%%%%%%%%%%%%%%%%%%%%%%%%%%%
\section{$H_1^\perp$ from $e^+e^-$ Annihilation ($e^+e^-\to h_1h_2X$)}
Recently we proposed a way to obtain information on the first 
$\bm{k}_T^2$-moment of 
$H_1^\perp$ from a $\cos(2\phi)$ angular dependence in two-hadron
production in $e^+e^-$ annihilation.~\cite{boe-97b} The two hadrons have to be
in the quark and anti-quark jet of an almost back-to-back jet event,
respectively, and their momenta should be reconstructed. Since the
fragmentation functions involved are $H_1^\perp$ and $\overline H_1^\perp$ (the
corresponding one for the anti-quark) the effect is independent of the spin of
the produced hadrons; to look for a pair of pions is the most obvious
choice, since they are abundantly produced.\footnote{This is spin physics with
spin-0 hadrons !} 
\begin{figure}[h]
\begin{center}
%\fbox{\begin{picture}(116,38) 
%\put(25,-1)
{\epsfig{figure=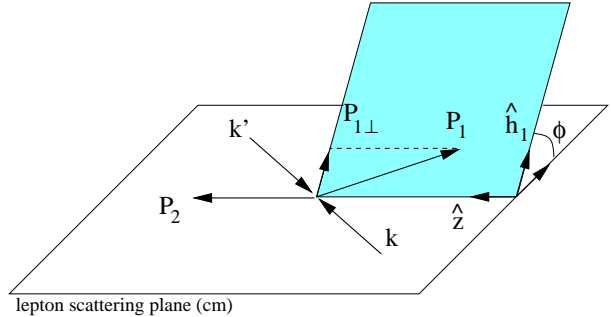,width=80mm}}
%\end{picture}
%}
\caption{Kinematics for two-hadron production in $e^+e^-$ annihilation.
} 
\end{center} 
\end{figure}
Again the appropriate way to pick out the $\cos(2\phi)$-dependent part of the
differential cross section is a weighted cross section
\begin{equation}
\left\langle W\right\rangle\;
\stackrel{\mbox{def}}{=}\;
\int\,\frac{d\phi^\ell}{2\pi}\,d^2{\bf q}_T\,W\,
\frac{d\sigma(e^+e^-\to h_1h_2X)}{d\Omega\,dz_1\,dz_2\,d^2{\bf q}_T}.
\end{equation}
At the Z-pole, i.e., $Q^2\simeq M_Z^2$, where most of the LEP data 
are available, the annihilation via a $Z$ boson dominates and the product of 
$H_1^{\perp(1)a}$ with $\overline{H}_1^{\perp(1)a}$ can be obtained from the
weighted cross section\footnote{Strictly, this gives information for the
quantities only at the scale of the $Z$ mass; evolution has to be taken into
account for usage in other hard processes at a different scale.}
\begin{equation} 
\left\langle\frac{Q_T^2}{4M_1M_2}\cos(2\phi)\right\rangle=
\frac{3\alpha^2\,Q^2\,y(1-y)\,
\sum_{a,\bar a}\,c_1^\ell c_2^a \;
H_1^{\perp(1)a}(z_1)\,\overline{H}_1^{\perp(1)a}(z_2)}
{4\,s_W^2\,c_W^2\left[(Q^2-M_Z^2)^2+\Lambda_Z^2M_Z^2\right]}
\end{equation}
where $s_W$($c_W$) is the $\sin$($\cos$) of $\theta_W$; the 
factors $c_1=g_V^{2}+g_A^{2}$, 
$c_2=g_V^{2}-g_A^{2}$ are relevant combinations of
(axial-)vector couplings (for later use: $c_3=2\,g_A\,g_V$).
%
%%%%%%%%%%%%%%%%%%%%%%%%%%%%%%%%%%%%%%%%%%%%%%%
\section{Electroweak $\ell H\to\ell^\prime hX$}
During this conference we have heard that high energy experiments reach the
precision where effects from electroweak currents have to be 
considered~\cite{EWremark}. We have calculated the differential cross section
for one-hadron inclusive lepton-hadron scattering thereby taking  
into account\\[-5.4mm]
\begin{itemize} 
\item the exchange of a photon or a $Z$ boson 
(including interference terms)\\[-5.4mm]
\item polarized beam/target/final state hadron (spin-1/2)\\[-5.4mm]
\item dependence on the transverse momentum of the produced hadron.\\[-5mm]
\end{itemize}
The full results will be presented elsewhere~\cite{boe-98b}. Here we give some
examples in form of weighted cross sections listed in Table 1. The 
electromagnetic part in the cross section of
the first line is the asymmetry discussed in Sec.~2. Additionally, from the
interference term there is a $\cos(\phi_h^\ell+\phi_S^\ell)$ dependence
involving the same combination of distribution and fragmentation functions 
(second line). In principle, it provides an independent piece of information 
for a flavor 
decomposition. But since, the term is probably difficult to measure 
accurately, one may think of a more modest usage. Together with the 
simplifying assumption of 
only one flavor contributing dominantly to the fragmentation, e.g. $u$-quark
fragmenting to a $\pi^+$ in the region of large $\xb$ and large $z_h$, this 
may be used for an independent cross-check on $h_1(x)\,H_1^{\perp(1)}$. 

The electromagnetic part of the cross-section in the third line was 
discussed before~\cite{kot-95}. Also here we find an orthogonal angular
dependence resulting from the interference of electromagnetic and weak
interaction (fourth line).

Particularly interesting is the possibility of non-zero ``naive
time-reversal odd'' distribution functions, which would show up in a weighted
cross section like in the fifth line. The electromagnetic term was discussed
by Boer and Mulders~\cite{boe-98a}.

\newpage
\begin{table}[ht]
\begin{center} 
\begin{tabular}{c|@{$\quad$}l}
\hline\hline
\rule[-3.6mm]{0pt}{8.8mm}$W$ & 
$\langle W\rangle
 \cdot\left[-\frac{4\pi\alpha^2s}{Q^4}\,|\bm{S}_T|\,(1-y)\,\xb\right]^{-1}$\\
\hline\\[-3mm]
$\frac{Q_T}{M_h}\sin(\phi_h^\ell+\phi_S^\ell)$ &
$\sum_{a,\bar a}\,\left(e_a^2+8\,g_V^\ell\,e_a\,g_V^a\,\chi_1
 +16\,c_1^\ell\,c_2^a\,\chi_2\right)$\\
&\qquad$\times\;h_1^a(\xb)\,H_1^{\perp(1)a}(z_h)$\\[1.2mm]
\hline\\[-3mm]
$\frac{Q_T}{M_h}\cos(\phi_h^\ell+\phi_S^\ell)$ &
$\sum_{a,\bar a}\,8g_V^\ell\,e_a\,g_A^a\,\chi_1\,
 \frac{-\Gamma_Z M_Z}{Q^2-M_Z^2}
\;h_1^a(\xb)\,H_1^{\perp(1)a}(z_h)$\\[1.2mm]
\hline\\[-3mm]
$\frac{Q_T^3}{6M^2M_h}\sin(3\phi_h^\ell-\phi_S^\ell)$ &
$\sum_{a,\bar a}\,\left(e_a^2+8\,g_V^\ell\,e_a\,g_V^a\,\chi_1
                             +16\,c_1^\ell\,c_2^a\,\chi_2\right)$\\
&\qquad$\times h_{1T}^{\perp(2)a}(\xb)\,H_1^{\perp(1)a}(z_h)$\\[1.2mm]
\hline\\[-3mm]
$\frac{Q_T^3}{6M^2M_h}\cos(3\phi_h^\ell-\phi_S^\ell)$ &
$\sum_{a,\bar a}\,8\,g_V^\ell\,e_a\,g_A^a\,\chi_1
\;h_{1T}^{\perp(2)a}(\xb)\,H_1^{\perp(1)a}(z_h)$\\[1.2mm]
\hline\\[-3.6mm]
$\frac{Q_T}{M_h}\sin(\phi_h^\ell-\phi_S^\ell)$ &
$\sum_{a,\bar a}\,
\bigg(\frac{2-2y+y^2}{2-2y}\,
\left(e_a^2+8\,g_V^\ell\,e_a\,g_V^a\,\chi_1+16\,c_1^\ell\,c_1^a\,\chi_2
\right)$\\
& \hspace{-12pt} $+\frac{y(2-y)}{2-2y}
\left(8 g_A^\ell e_a g_A^a \chi_1+16 c_3^\ell c_3^a \chi_2
\right)\!\!\!\bigg)
f_{1T}^{\perp(1)a}(\xb) D_1^a(z_h)$\\
\hline\hline
\end{tabular}\\[1mm]
\begin{tabular}{c@{\qquad\qquad}c}
$\chi_1=\frac{1}{16\,s_W^2\,c_W^2}\;
\frac{Q^2(Q^2-M_Z^2)}{(Q^2-M_Z^2)^2+\Gamma_Z^2M_Z^2}$ &
$\chi_2=\frac{1}{16\,s_W^2\,c_W^2}\;\frac{Q^2}{Q^2-M_Z^2}\;\chi_1$
\end{tabular}
\caption{Weighted cross sections in semi-inclusive DIS.}
\end{center} 
\end{table}
\vspace*{-7.2mm}
\section*{Acknowledgments}
\vspace*{-2.6mm}
{\footnotesize This work is part of the research program of the 
foundation for Fundamental Research of\\[-1.2mm] 
Matter (FOM), the National Organization for Scientific Research (NWO)
and the TMR\\[-1.2mm] 
program ERB FMRX-CT96-0008.}

\end{document}